\documentclass[aps,prl,twocolumn,showkeys]{revtex4-1}

\usepackage[cmex10]{amsmath}
\usepackage{graphicx,epstopdf}
\usepackage{amssymb}
\usepackage{algorithmic}
\usepackage[linesnumbered,ruled]{algorithm2e} 

\newcommand{\DTW}{\text{DTW}}
\newcommand{\LBK}{\text{LBK}}
\newcommand{\vect}[1]{\mathbf{#1}}
\newcommand{\reffig}[1]{Fig.~\ref{#1}}
\newcommand{\refeq}[1]{Eq.~\ref{#1}}
\newcommand{\reftbl}[1]{Table~\ref{#1}}
\newcommand{\refsec}[1]{Sec.~\ref{#1}}

\usepackage[iso]{datetime}
\usepackage[colorlinks=true,linkcolor=red,urlcolor=blue,citecolor=red]%
	{hyperref}
\begin{document}
\title{
%	{\normalsize \hbTimeStamp}\\
	An Accelerometer Based Calculator for 
	Visually Impaired People Using Mobile Devices 
}

\author{
	Dogukan~Erenel and 
	Haluk~O.~Bingol
}
\affiliation{The Department of Computer Engineering, 
	Bogazici University, 
	Istanbul
}

\begin{abstract}
	Recent trend of touch-screen devices produces an accessibility barrier 
	for visually impaired people.
	On the other hand, these devices come with sensors such as accelerometer.
	This calls for new approaches to human computer interface (HCI).
	In this study, our aim is to find an alternative approach to 
	classify 20 different hand gestures captured by 
	iPhone 3GS's built-in accelerometer and 
	make high accuracy on user-independent classifications using 
	Dynamic Time Warping (DTW) with dynamic warping window sizes.
	20 gestures with 1,100 gesture data are collected 
	from 15 normal-visioned people.
	This data set is used for training.
	Experiment-1 based on this data set produced an accuracy rate of 96.7~\%.
	In order for visually impaired people to use the system,
	a gesture recognition based ``talking'' calculator is implemented.
	In Experiment-2, 4 visually impaired end-users used the calculator
	and obtained 95.5~\% accuracy rate 
	among 17 gestures with 720 gesture data totally.
	Contributions of the techniques to the end result is also investigated.
	Dynamic warping window size is found to be the most effective one.
	The data and the code is available.
\end{abstract}


% =========================== 
\maketitle

% =========================== 
\section{Introduction}

Within the popularity of new devices 
such as accelerometer based game controllers or touch-screen smartphones, 
the need of new human computer interfaces emerged.
This is especially true in the area of accessibility although
some mobile devices with just touch-screens come with features 
such as text-to-speech, speech-to-text, magnifier for handicapped people.
Several research works on accelerometer based gesture recognition systems 
and on the usage of accelerometer based devices 
in medical area pioneered new interfaces for accessibility. 
For example, the Nintendo Wii controller is used for patients recovering from 
strokes, broken bones, surgery and even combat injuries with 
some specific 
games~\cite{
	Deutsch2008,
	Kratz2007, 
	Schlomer2008}. 

There are limited research works on these mentioned interfaces 
for visually impaired people on mobile devices. 
Text editing on a touch-screen device is one of the major issue.
Clearly, text-to-speech, speech-to-text systems would be ultimate solutions.
Handwriting on the screen is an other one.
On the other hand, 
accelerometer based systems are also a potential candidate at least for some domains.
This work aims to present a solution to this problem 
for the limited domain of arithmetic calculations.

Several methods have been suggested with 
different approaches for an accelerometer based gesture recognition system, 
which are mostly used 
\emph{Hidden Markov Models 
(HMM)}~\cite{
	Mantyla2000, 
	Mantyjarvi2004,
	Kela2006}.
Some of them are applied on mobile devices, for example; 
Pylvanainen proposed a gesture recognition system based on 
continuous 
HMM~\cite{
	Pylvanainen2005}. 
Prekopcsak uses HMMs and \emph{Support Vector Machines (SVM)} to classify 
gestures captured by built-in accelerometer of a mobile phone, 
namely Sony-Ericsson W910i~\cite{
	Prekopcsak2008}.
In addition, Klingmann uses HMMs with iPhone built-in 
accelerometer~\cite{
	Klingmann2009}.
As an alternative solution to HMM, 
Wu et al. proposes an acceleration-based 
gesture recognition approach using SVM with a Nintendo Wii 
controller~\cite{
	Wu2009}.
Besides using HMM or any probabilistic approaches, 
some researches represents 
\emph{Dynamic Time Warping (DTW)} with template adaptation. 
For example; uWave includes quantization of accelerometer readings, 
DTW and template adaptation using a mobile 
device~\cite{
	Liu2009PMC,
	Liu2009}.
Leong et al. uses DTW with a Nintendo Wii 
controller~\cite{
	Leong2009}.
Akl and Valaee use DTW 
as well as affinity propagation with a Nintendo Wii 
controller~\cite{
	Akl2010,
	Akl2011}.

In this work, a reliable, fast and simple 
gesture recognition model
and its implementation as a new 
interface is developed.
The model is based on 
the technique originally proposed in Ratanamahatana and Keogh's 
work~\cite{
	Ratanamahatana2004, 
	Erenel2011}.
As a proof of concept, a simple ``talking'' calculator is implemented. 
Among the main contributions of this work are
%- 
a new interface to 
write text by capturing accelerometer data of hand gestures for 
touch-screen smartphones with built-in accelerometer
%-
and a detailed analysis of the contributions of techniques that are used.
The proposed system is capable of classifying 20 different gestures 
with high reliability.
The system has been tested by visually impaired end-users with 
the implemented application on iPhone 3GS.

% =========================== 
\section{Method}

An accelerometer based gesture recognition system is proposed which
consists of three parts, namely,
data collection, training and classification.
Hand gesture data from participants are collected in data collection part using iPhone.
Then all captured data are processed and a classifier is trained and validated 
in training part using a desktop computer.
Finally, the trained classifier is tested by visually impaired participants in classification part via iPhone.

% =========================== 
\subsection{Data Collection}

% =========================== 
\subsubsection{Gesture Set}

% ++++++++++++++++++++++++++++++++++++++++++++++++++++++++++++++
\begin{figure}
	\centering
	\includegraphics[width=.90\columnwidth]%
		{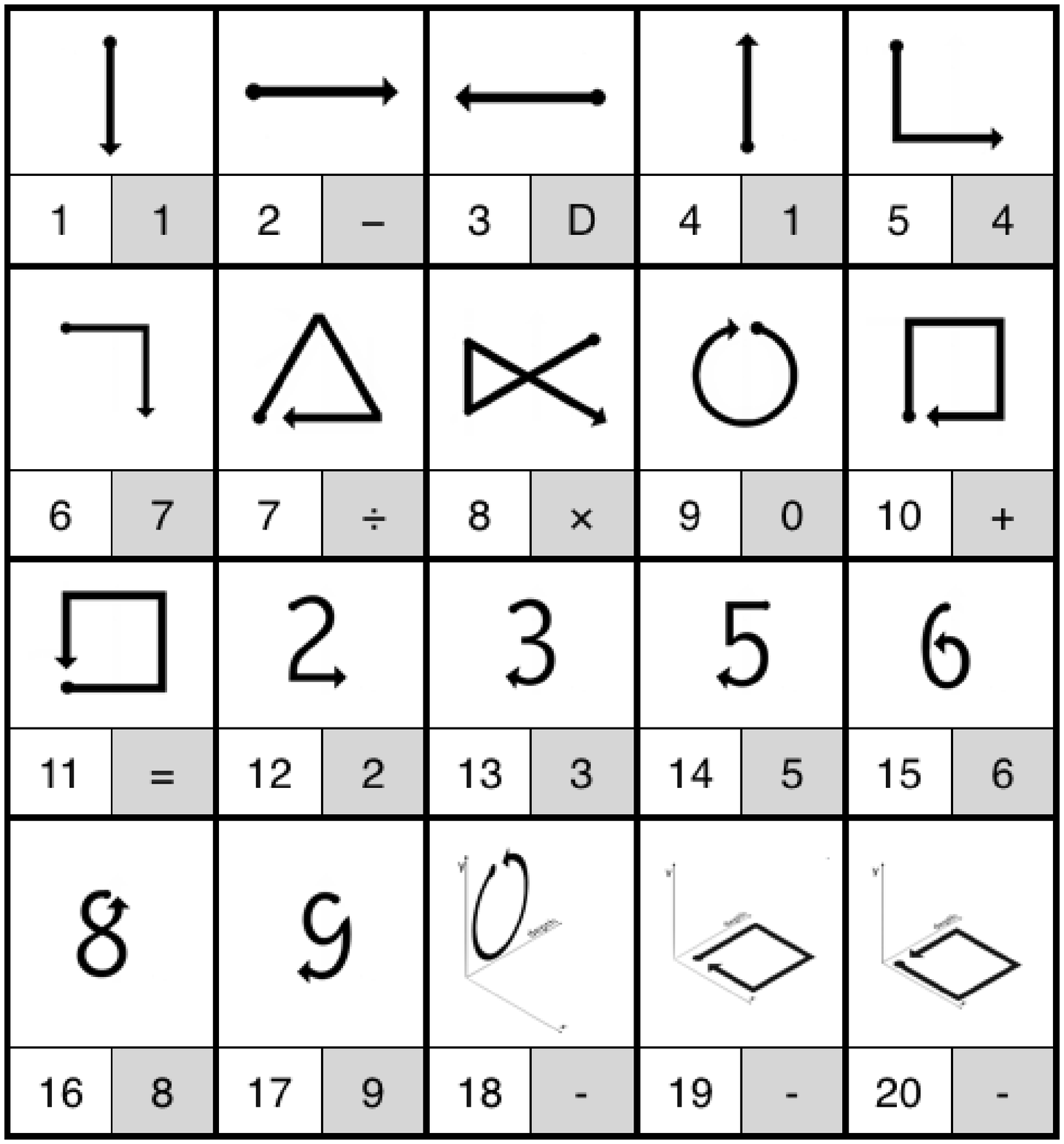}
	\caption[Gesture Set]{
		Gesture Set with 20 gestures.
		The gesture ID is given in the lower left white box of the gesture.
		Gestures 1-4 are in 1D,
		5-17 in 2D, and
		18-20 in in different plane than 5-17.
		The interpretation of the gestures 
		in the calculator are given in the lower right grey box.
		Gestures 1 and 4 correspond to digit 1.
		All the other symbols have exactly one corresponding gesture.
		Gestures 10, 2, 8, 7, 11 and 3 are mapped to 
		$+, -, \times, \div, =$ and ``delete-the-last-entry'', respectively. 
		Gestures 18, 19 are 20 are not used in the calculator.
	}
	\label{fig:gestureSet20}
\end{figure}
% ++++++++++++++++++++++++++++++++++++++++++++++++++++++++++++++

20 gestures, given in \reffig{fig:gestureSet20}, are designed.
There are two design criteria. 
(i)~The gestures should be intuitive so that they can be remembered easily.
(ii)~While doing a gesture,
no visual clues should be necessary so that visually impaired could do it.
Since a calculator is in mind, 
gestures corresponding to digits and arithmetic operations are necessary.
Gestures very similar to shape of digits $0-9$ are used. 
Gestures for $-$, $\times$ are also similar to their shapes in mathematics.
Gesture for ``delete-the-last-entry'' reminds erasing. 
On the other hand, 
gestures for $+$, $\div$, and $=$ are not that intuitive.

Note that the gestures are in different dimensions.
Gestures 1-4 are in 1D, only.
Gestures 5-17 are in 2D. 
The remaining 3 gestures, 18-20, are in a plain different then that of 5-17.

% =========================== 
\subsubsection{Device and Data Representation}

iPhone 3GS is used as the device for data collection.
It's built-in accelerometer measures the proper acceleration 
which is the sum of accelerations due to 
gravitation and the gesture motion.
The unit of measurements is in terms of $g$
where $g$ is the gravitational acceleration due to the Earth.
It has a range of $\pm 2g$
and a sensitivity of approximately $0.02g$ 
If the phone is laying on its back on a horizontal surface, 
acceleration values (in 3D) will be approximately the following values: 
$x = 0, y = 0, z = -1$, all in $g$. 
 
The accelerometer,
which is configured to capture data at 60 Hz,
produces four time series: 
three for each axes, namely, $x(t), y(t), z(t)$,  
and one for the 
time~\cite{%
	St2008}. 
A sequence of acceleration vectors sampled at discrete times 
$k = 1, 2, \dotsc, K$ is represented as 
\[
	\vect{A} = \{ \vect{a}(k) \}_{k=1}^{K}
\]
where
$\vect{a}(k) \triangleq [ x(k), y(k), z(k) ]^{\top}$ 
is an 3D column vector at time step $k$.
Note that $\{ \vect{a}(k) \}_{k=1}^{K}$ is a 3D signal.
1D signals 
$\{ x(k) \}_{k=1}^{K}$,
$\{ y(k) \}_{k=1}^{K}$,
$\{ z(k) \}_{k=1}^{K}$,
are called \emph{channels}.

% =========================== 
\subsubsection{Mobile Applications}

One iPhone application with multi views is developed.
The data acquisition view is used to collect acceleration data
while user does gestures.
User is asked to make the gesture while phone is facing her.
She presses a finger on the screen to start data collection. 
Data is kept collected as long as the finger is on the screen. 
It stops when the finger is removed from the screen.

The talking calculator view is 
a simple ``talking'' integer calculator with 4-operation 
which is used for testing our approach by visually impaired users
who needs audio feedback.
An 4-operation calculator requires 16 different symbols 
(10 for digits, 4 for operations, one for ``='' and 
one for ``delete-the-last-entry'').
Based on familiarity to the symbols, 
17 gestures from \reffig{fig:gestureSet20} are selected for the calculator.
Note that digit 1 has tow corresponding gestures, namely 1 and 4.
Gestures 18, 19 are 20 are not used in the calculator.
Text-to-speech library ``Flite''~\cite{urlFlite}
and its wrapper by Sam Foster
is used to ``speak'' of the gesture that is 
entered~\cite{
	Black2001%, 
}. 
The code is available at \cite{urlDataCode}

% =========================== 
\subsubsection{Users and Data Acquisition}

The gesture data set is collected from 15 users.
Users are undergraduate students, mainly freshmen and sophomores, of our department.
It is necessary to point out that 
since they are Computer Engineering students,
they may be more familiar to this then an average person.

We want user to be in their every day environment.
There were no particular time and place for data collection.
We asked students to participate during the break between courses.
There were no problem about usage of the application or the gesture set that is reported.

We show a user how we place our finger on the screen and do the gesture.
This done once and no further training is given.
Then we give the phone to the user and 
she makes the gestures in \reffig{fig:gestureSet20}.
 
Each user is asked to do 20 gestures multiple times, 
so on the average 55 gesture data are collected for each gesture.
This makes in total 1,100 gesture data,
out of which 10 recordings are found to be faulty.
These 10 recordings were too short to process.
Hence 1,090 gesture data is used in this study.
The data is available at \cite{urlDataCode}

% =========================== 
\subsection{Training}

After data set generation at data collection, training processes are taken place.
Training is computational intensive task,
which is done on computer.
Once system learns to recognize the gestures,
then the trained system is moved to mobile device.

\begin{figure*}
	\centering
	\includegraphics[width=1.6\columnwidth]{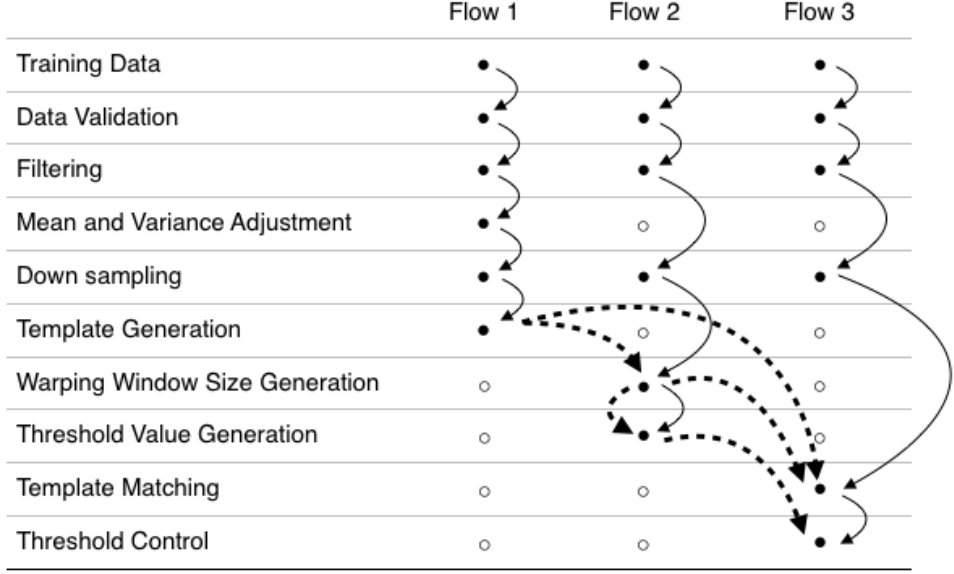}
	\caption[Processing Blocks]{
		The Flow-1 produces templates $\vect{G}_{m}$, 
		lower and upper bounds 
		$\vect{L}_{m}$ and 
		$\vect{U}_{m}$,
		respectively.
		Then, the Flow-2 generates the warping window sizes 
		$\vect{W}_{m}$, 
		threshold values 
		$D^{\min}_{m}$ and
		$D^{\max}_{m}$.
		The Flow-3 represents the classification flow 
		which uses the values generated in the first two flows.
	}
	\label{fig:ProcessingBlocks}
\end{figure*}

Overall system is given in \reffig{fig:ProcessingBlocks} as block diagrams,
which will be considered later in \refsec{sec:ProcessingBlocks}. 
Flow-1 produces the gesture templates.
In Flow-1,
the training raw data which contains 1,090 gesture data,
is processed by 
	validation, 
	low-pass filtering,
	mean and variance adjustment,
	down sampling, 
	template generation
operations.

In Flow-2 in \reffig{fig:ProcessingBlocks}, 
the training raw data
is processed by means of 
	validation, 
	low-pass filter,
	down sampling, 
	warping window size generation,
	threshold values generation.
Warping windows and threshold values of corresponding gesture representatives 
are obtained as a result of Flow-2.

% ++++++++++++++++++++++++++++++++++++++++++++++++++++++++++++++
% ++++++++++++++++++++++++++++++++++++++++++++++++++++++++++++++

% =========================== 
\subsection{Classification}

The gesture done by user needs to be classified on a mobile device, in our case iPhone.
In Flow-3 in \reffig{fig:ProcessingBlocks} is the classification 
which passes through the following processes:
	validation, 
	low-pass filter,
	mean and variance adjustment,
	down sampling,
	template matching,
	threshold control.
Then the system gives the output as classification result.
The gesture is classified using template matching algorithm 
and the closest valid gesture representative is given as classification result.

% =========================== 
\section{Processing Blocks}
\label{sec:ProcessingBlocks}

The 3D raw gesture data 
$\vect{A} = \{ \vect{a}(k) \}_{k=1}^{K}$ 
collected from user is passed through a number of processing blocks
schematically given in \reffig{fig:ProcessingBlocks}.

% ===========================
\subsection{Validation}

Clearly, every user has her own paste of doing a gesture. 
Some does the gesture fast, some does it slow.
Similarly, some user does the same gesture in a small scale, 
some in a large scale.
That effects the duration of gesture data.
We discard gesture data that is 
too short ($K < K_{\min}$) or 
too long ($K_{\max} < K$) in duration.
We use 
$K_{\min} = 30$ and 
$K_{\max} = 205$.

Second validation is related to the amplitude.
It is expected that 
the amplitude of the signal changes as user draws the gestures.
We restrict the average amplitude in a acceptable range.
Since our gesture data is in 3D,
the average amplitude of a gesture 
$\vect{A} = \{ \vect{a}(k) \}_{k=1}^{K}$ is defined as 
$a_{\text{avg}} = \frac{1}{K} \sum_{k=1}^{K} |\!| \vect{a}(k) |\!|$
where 
$|\!| \vect{a}(k) |\!|$ is the magnitude of 
$\vect{a}(k)$.
Data sets with average amplitude 
too small ($a_{\text{avg}} < R_{\min}$) or 
too big ($R_{\max} < a_{\text{avg}}$) 
are also discarded.
We use 
$R_{\min} = 0.95$ and 
$R_{\max} = 2.10$.

Out of 1,090 data sets, 
24 due to duration and 4 due to amplitude, 
in total 28 are discarded and 
we end up with 1,062 gesture data for 20 gesture classes.

% ===========================
\subsection{Filtering}

The high frequency components in each channel 
are removed by means of a low-pass filter given as
$y_{k} = \alpha x_{k} + (1 - \alpha) y_{k-1}$
where
$x$ and $y$ are the input and the output signals of the filter, respectively, and
the smoothing factor taking to be $\alpha=1/7$.

% ===========================
\subsection{Adjustment of Mean and Variance}

Every gesture is different, 
hence it has different characteristics.
After filtering,
we adjust the mean and variance of 
data for each gesture individually
so that each gesture has its own average and variance.

Since our gesture data is in 3D,
we apply mean and variance adjustments to every channel individually.
We obtain zero-average channel signal 
by subtracting the average of the channel.
Then we get the \emph{mean adjusted channel}
by adding the average of the channel 
over all the signals of gesture $m$.

For variance adjustment,
we obtain the variance of the channel.
Then we get the average of all the variances of the channel 
over all the signals of gesture $m$.
Finally each gesture data for $m$ are adjusted so that 
each channels share the same mean and the variance of the gesture.

% ===========================
\subsection{Down Sampling}

So far each gesture data has different duration. 
We down sample each gesture data in such a way that 
they have the same durations of $N$.
We use $N = K_{\min}$,
which is the acceptable minimum duration.

If the mean and variance adjusted gesture data 
$\vect{A} = \{ \vect{a}(k) \}_{k=1}^{K}$
has $K$ data points,
we need to use downsampling factor of $\Delta = K/N$.
That is,
we represent 
every consecutive $\Delta$ data points with one data point.
We obtain the downsampled data 
$\vect{D} = \{ \vect{d}(n) \}_{n=1}^{N}$
using the following averaging
\[
	\vect{d}_{j}(n) 
	\triangleq 
	\frac{1}{\Delta} 
	\sum_{\substack{k \\ (n-1) \Delta < k \leq n \Delta}} 
		\vect{a}_{j}(k).
\]

% ===========================
\subsection{Templates}

For each gesture $m$, 
we want to generate a template 
$\vect{G}_{m}$ so that 
a given gesture data 
$\vect{X}$ is classified to 
class $m_{j}$ 
if $\vect{X}$ is closest to $\vect{G}_{m_{j}}$ 
with respect to a distance metric.
We simple consider the average of all the gesture data of the gesture $m$ 
as its template.

DTW is used as the distance metric.
In order to speed up,
the $\LBK$ technique is employed 
which requires 
lower $\vect{L}_{m}$ and 
upper $\vect{U}_{m}$ 
bounds for each gesture class $m$.
Template generation also produces 
$\vect{L}_{m}$ and 
$\vect{U}_{m}$ in two steps:
(i)~The lower bound $\vect{L}_{j}$ and upper bound $\vect{U}_{j}$ of 
gesture data $\vect{A}_{j}$ in the gesture set of $m$ is calculated 
for each channel individually as given in \cite{Keogh2004} 
using $\LBK$ parameter $r = 3$.
(ii)~Then, the bounds for gesture $m$ is obtained by averaging
lower $\vect{L}_{j}$ and 
upper $\vect{U}_{j}$ bounds of each gesture data obtained in step (i).

% =========================== 
\subsection{Warping Window Size}

For each gesture class $m$, 
a specific sequence of warping window sizes 
$\vect{W}_{m} = \{ \vect{w}_{m}(n) \}_{n=1}^{N}$ 
is generated where $\vect{w}_{m}(n)$ is the window size at time $n$.
Warping window size generation is based on Ratanamahatana and Keogh's 
work~\cite{Keogh2004}. 
The warping window size $w(n)$ minimizes  
the quality metric $Q$ of \cite{Ratanamahatana2004}.
That is,
\[
	w(n) = \arg \min_{w} \{ Q \}
\]
at each step $n \in \{ 1, 2, \dotsc, N \}$.

% ===========================
\subsection{Threshold Values}

Distance of 
gesture data $\vect{A}_{j}$ for gesture $m$
to the template $\vect{G}_{m}$ is given as 
$\DTW( \vect{A}_{j}, \vect{G}_{m}, \vect{W}_{m})$.
We want to control the distance to the template
by the minimum and maximum 
of these distances are given as
\[
	D^{\min}_{m}
	\triangleq 
	(1 - K_{\text{D}})
	\min_{\vect{A}_{j} \in \mathcal{A}_{m}}
	\{ 
		\DTW( \vect{A}_{j}, \vect{G}_{m}, \vect{W}_{m}) 
	\} 
\]
and
\[
	D^{\max}_{m}
	= 
	(1 + K_{\text{D}})
	\max_{\vect{A}_{j} \in \mathcal{A}_{m}}
	\{ 
		\DTW( \vect{A}_{j}, \vect{G}_{m}, \vect{W}_{m})
	\}, 
\]
respectively,
where 
$\mathcal{A}_{m}$ is the set of all gesture data for $m$, and 
$K_{\text{D}}$ is a safety constant taken to be $K_{\text{D}} = 0.1$.

% ===========================
\subsection{DTW Template Matching}

Gesture $\vect{A}_{j}$ is classified to gesture class $m_{c}$ if 
$\DTW( \vect{A}_{j}$, $\vect{G}_{m_{c}}$, $\vect{W}_{m_{c}}) $ 
is the smallest for all $m$.
That is,
\[
	m_{c} 
	= \arg \min_{m} 
	\{ 
		\DTW( \vect{A}_{j}, \vect{G}_{m}, \vect{W}_{m} ) 
	\}.
\]

This calls for repeated evaluation of 
$\DTW(\vect{A}_{j}, \vect{G}_{m}$, $\vect{W}_{m})$ for each $m$. 
The evaluation is speeded up by means of $\LBK$ technique 
using $L_{m}$ and $U_{m}$ generated in the template generation.

% ===========================
\subsection{Threshold Control}

Threshold values generated previously for given gesture class is used 
for classification result validation.
If 
$D^{\min}_{m_{c}} 
< \DTW(\vect{A}_{j}, \vect{G}_{m_{c}}, \vect{W}_{m_{c}}) 
< D^{\max}_{m_{c}} $,
then $m_{c}$ is the valid classification result. 
Otherwise, $m_{c}$ is discarded and classification result is invalid.

% =========================== 
\section{Experiments and Results}

There are two experiments in this study.

% ++++++++++++++++++++++++++++++++++++++++++++++++++++++++++++++
\begin{figure}%[here]
	\centering
	\includegraphics[width=1.1\columnwidth]%
		{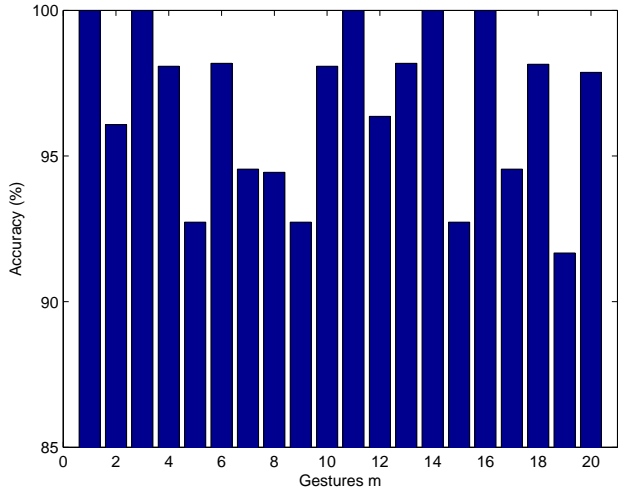}
	\caption[Experimental Result 1]{
		Recognition accuracy rate for each gesture class 
		obtained in Experiment-1.
	}
	\label{fig:AccuracyByGesture}
\end{figure}
% ++++++++++++++++++++++++++++++++++++++++++++++++++++++++++++++

% =========================== 
\subsection{Experiments-1}

Experiment-1 is the system validation test.
The data collected from normal users 
for template generation is used in Experiment-1.
In Experiment-1, system is trained with training data set.
Then it is validated with collected data using 
Flow-2 given in \reffig{fig:ProcessingBlocks}.
There are 1,090 gesture data in validation set.  
The average classification accuracy is 96.7~\%.
In addition, the recognition accuracy for each gesture class is 
given in \reffig{fig:AccuracyByGesture}.

% =========================== 
\subsection{Experiments-2}

% ++++++++++++++++++++++++++++++++++++++++++++++++++++++++++++++
\begin{table}%[h]
	\caption{All 40 calculations used in Experiment-2.}
	\centering
	\begin{tabular}{ | l | l | l | l | }
		\hline
		$6+9=$ &$72-4=$ &$3\times8=$ &$1\div50=$\\ \hline
		$7+9=$ &$1-24=$ &$8\times6=$ &$30\div5=$\\ \hline  
		$5+1=$ &$8-9=$ &$76\times3=$ &$40\div2=$\\ \hline
		$90+7=$ &$3-6=$ &$1\times5=$ &$8\div24=$\\ \hline  
		$8+3=$ &$56-2=$ &$7\times9=$ &$10\div4=$\\ \hline 
		$2+0=$ &$6-7=$ &$31\times4=$ &$58\div9=$\\ \hline
		$4+5=$ &$8-7=$ &$90\times2=$ &$6\div13=$\\ \hline 
		$67+4=$ &$1-9=$ &$3\times80=$ &$5\div2=$\\ \hline
		$30+4=$ &$1-5=$ &$67\times9=$ &$2\div8=$\\ \hline 
		$81+9=$ &$30-5=$ &$4\times7=$ &$2\div6=$\\ \hline
	\end{tabular}
	\label{tbl:calculations}
\end{table}
% ++++++++++++++++++++++++++++++++++++++++++++++++++++++++++++++

Once templates are generated using training data from normal users,
performance of the method for the actual target users is investigated.
Experiment-2 is the end-user test.
Visually impaired users use the calculator to perform some calculations.
	
For Experiment-2, a test set of 40 calculations, 
given in \reftbl{tbl:calculations}, is designed.
In order to do all the calculations user has to enter 180 characters in total.
Note that each row of the table contains 
10 digits, 
4 operators and 
4 ``='' symbols.
Therefore each symbol, except ``='', has to be entered 
10 times during a test.

Experiment-2 is performed by 4 visually impaired participants, 
who did not attend in gesture data collection.
A demo video of the application usage by a visually impaired participant is 
available on the web~\cite{urlVideo}
%%	KEEP
%%   http://www.youtube.com/watch?v=VPzg3gHINPs
%%

\begin{figure}[here]
	\centering
	\includegraphics[width=1.1\columnwidth]%
		{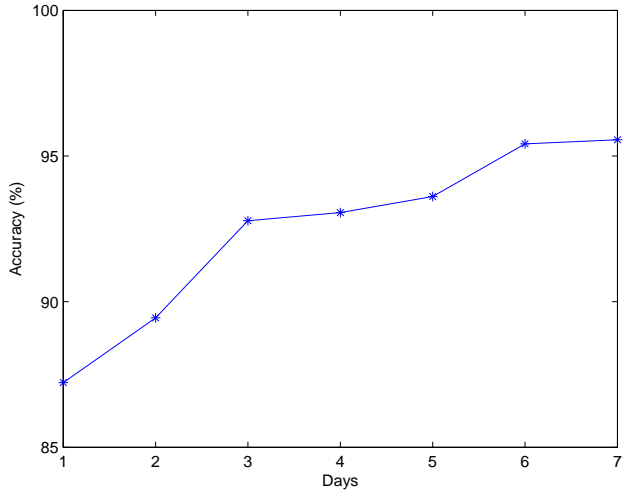}
	\caption[Experimental Result 2]{%
		Daily average recognition accuracy rate
		obtained from Experiment-2. 
	}
	\label{fig:AccuracyByDay}
\end{figure}

We want to investigate not only one time usage 
but also adaptation of users to the system.
The test lasts for an adaptation period of 7 days.
In the first day, each participant is trained 10 minutes about 
the system, the gesture movements and their meanings, 
the phone holding position and voice feedback.
Then, each participant
did the 40 calculations once a day for 7 days.
The average daily recognition accuracy is given in \reffig{fig:AccuracyByDay}.
The average recognition accuracy increases day by day, 
to reach 95.5~\% in the seventh day.

% ++++++++++++++++++++++++++++++++++++++++++++++++++++++++++++++
% ++++++++++++++++++++++++++++++++++++++++++++++++++++++++++++++

% =========================== 
\section{Discussion}

% =========================== 
\subsection{Comparisons}

There are several research works related to proposed method.
In means of user-independent result; 
uWave gives 75.4~\% for 8 
gestures~\cite{%
	Liu2009}, 
Leong et al. founds 72~\% for 10 gestures~\cite{Leong2009}, 
Akl and Valaee give $\approx$ 90~\% for 18 gestures 
(among not-included users)~\cite{%
	Akl2010}.
Note that, users in Experiment-2 did not attend in data collection part 
and they are visually impaired.
Based on the user-independent classification accuracy rates, 
the number of gestures,
and end-user experiment; 
our proposed method seems to be one of the best among previous works.
There are a number of possible reasons for high accuracy:
(i)~We ask users to keep the phone facing them as much as possible while they are doing gesture.
This may reduce noise.
(ii)~User stars and stops the data collection by pressing a finger to screen.
So we get nothing but the gesture data.
(iii)~the subjects are Computer Engineering students
that are more suitable to such tests than general audience.

% ++++++++++++++++++++++++++++++++++++++++++++++++++++++++++++++
\begin{figure*}[ht]
	\centering
	\includegraphics[width=1.90\columnwidth]%
		{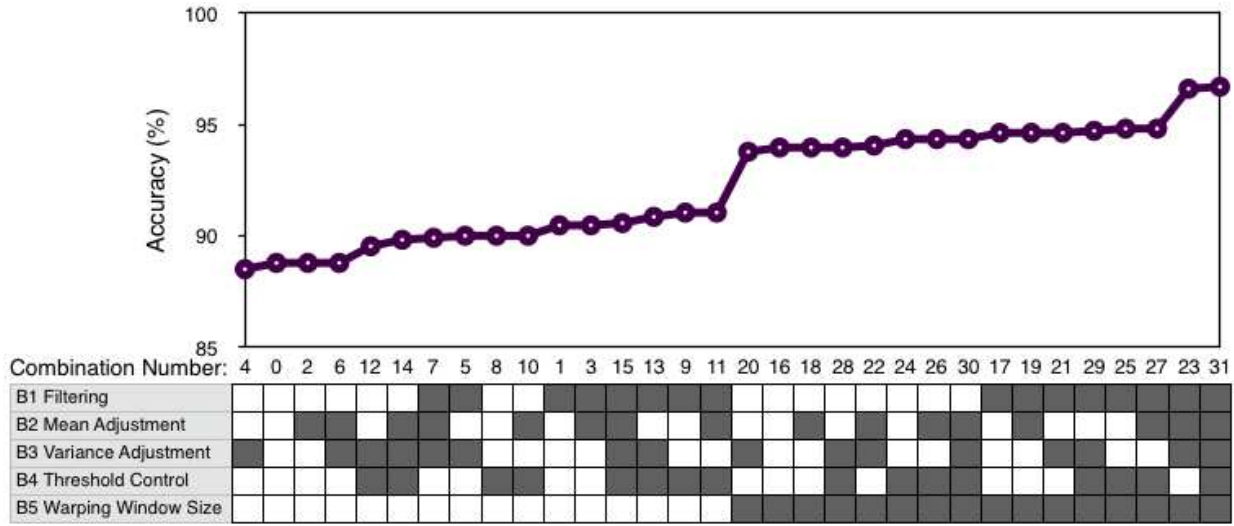}
	\caption{%
		Effects of some data processing blocks on classification accuracy.
		If the processing block is ``off'' the corresponding box is white, 
		if it is ``on'', the box is gray.
	}
	\label{fig:ProcedureEffects}
\end{figure*}
% ++++++++++++++++++++++++++++++++++++++++++++++++++++++++++++++

% =========================== 
\subsection{Contribution of the Blocks}

\newcommand{\hbBi}{B1 ``filtering''}
\newcommand{\hbBii}{B2 ``mean adjustment''}
\newcommand{\hbBiii}{B3 ``variance adjustment''}
\newcommand{\hbBiv}{B4 ``threshold control''}
\newcommand{\hbBv}{B5 ``warping window''}
We investigate the contribution of the processing blocks of
\refsec{sec:ProcessingBlocks} in various combinations.
One expects that each technique used
has different contributions to the classification accuracy.
The techniques are grouped into 5 data processing blocks.
The \emph{blocks} are; 
B1~filtering, 
B2~mean adjustment, 
B3~variance adjustment,
B4~threshold control,
and 
B5~using warping window size which includes template generation.
Note that, template matching and warping window size generation 
are considered as one block.
On the other hand, mean and variance adjustment are considered as 
two blocks separately.

\reffig{fig:ProcedureEffects} 
provides the accuracy obtained by 
all 32 possible combinations of these 5 blocks 
using the data of Experiment-1.
The combinations are ordered in the accuracy that they get.

One expect that adding a new block increases the accuracy 
but that is not the case.
The pattern is quite complex.
In some combinations,
adding a new block degrades the performance.
The very same block improves the performance 
if it is added to some other combination.
Block \hbBiii\ is one of them.
Out of 16 possible configurations of other four blocks,
adding B3 increases the performance
in only 5 of them. 
In the remaining 11, it decreases it.
See configuration pairs 
2-6, 
18-22 
for performance increase, and
pairs 
4-0,
13-9 for degrading.
 
It is noted that the effect of using block \hbBv\ is 
dramatic.
It is the primary reason of the step jump 
from the first 16 combinations on the left including 11, 
to the remaining 16 combinations on the right starting with 20 
in \reffig{fig:ProcedureEffects}.
Interestingly, 
just by itself,
it produces close to 95~\% accuracy
observed at combination 16.

Block \hbBi\ also has a big impact, too.
In the first 16 combinations without B5,
the highest 6 (from 1 through 11) includes B1.
In the second 16 combinations that have B5,
and 
the highest 8 combinations (from 17 through 31) uses B1.
Without any other blocks, only \hbBi\ and \hbBv\ together,
i.e., combination 17,
manage to obtain almost 95~\% accuracy .

% =========================== 
\subsection{Parameters}

A number of parameters
are used in the proposed system.
They are generally decided empirically.
Firstly, we decide to use iPhone built-in accelerometer at 
$F_{\text{sampling}} = 60\text{ Hz}$. 
Then we assume that a user makes a gesture movement 
between 0.5 to 3.5 second, 
which is equal 30 to 210 sample points at the given $F_{\text{sampling}}$.
We check the minimum and maximum lengths of our dataset 
and determine the value for $K_{\min} = 30$ and $K_{\max}= 205$. 
If we hold iPhone in a stable position, 
it's built-in accelerometer measures $1g$ as amplitude. 
A gesture is an accelerating movement that the start and 
stop values are known as $\approx 1 g$.
In addition, built-in accelerometer has a range of $\pm 2 g$ at each axis, 
which is same as $0g$ to $\approx 3.46 g$ in terms of amplitude.
After considering these conditions and an additional 5~\% for threshold value; 
we assumed that a gesture data has average amplitude between 
$R_{\min} =0.95 g$ to $R_{\max} =2.10 g$.
We use minimum gesture length as down sampling sample size $N$,
which is related to minimum number of samples $K_{\min}$. 
Finally, in threshold value generation we use $\pm 10~\%$ as 
a safety constant as $K_{\text{D}} = 0.1$.

% =========================== 
\subsection{Future work}

Lastly, one needs to consider points of improvement.
(i)~The main limitation is that the system is not working as 
an instant continuous recognition system.
User triggers the start and the stop of the gesture.
Making our proposed system to an continuous-gesture recognition system, 
which segments the data instantly and 
recognize the gesture, may be a good goal for future works.
(ii)~Our gesture set is selected among previous works and added some new ones.
This calls for research on design of gesture set 
which would improve the accuracy rate 
as well as usability of the system by the target end-users.  
(iii)~In order to improve accuracy some additional sensor, such as gyroscope, 
can be added to the system.
(iv)~Our proposed system generates gesture templates directly 
from training set.
The quality of training set directly effects the quality of the system.
How to measure and improve the quality of training set is an open issue 
yet to be investigated.
In our case the training set is collected from people 
with normal vision
but the end-user tests are done by visually impaired people.
It would be interesting to see the performance of the system 
when the training set is also collected from the end-users 
which we could not do due to lack of access to 
enough number of visually impaired people.

% =========================== 
\section{Conclusion}

Recent trent of touch-screen devices produces a barrier for 
visually impaired people.
This calls for new human-computer interfaces.
An optimized accelerometer based gesture recognition system is introduced
which hopefully contributes to the integration of the visually impaired 
to the society.
The system is designed on a touch-screen smartphones with 
built-in accelerometer, namely, iPhone 3GS.
The proposed method gives 96.7~\% accuracy on training set using 20 gestures.
As a proof of concept of the system, 
a gesture based simple calculator is implemented.
End-user test done by 4 visually impaired people,
who did not attend in data collection part,
using the calculator 
with 17 gestures obtains 95.5~\% accuracy.
In summary, our proposed gesture recognition system provides 
an outstanding performance in terms of user-independent recognition accuracy, 
experimental results of end-users and variety of gesture set 
when compared to other systems in literature.

A number of processing blocks are used.
Their contributions are investigated.
Interestingly, one block outperforms all.
That is, Warping Window Size has the largest impact to the end result.
No other individual block or a combination of blocks approach to 
the effect of it.

% =========================== 
\begin{acknowledgments}
This work is partially supported 
by the Turkish State Planning Organization (DPT) the TAM Project, 2007K120610 and
by Bogazici University Research Fund Project, BAP-2011-6017.
\end{acknowledgments}

\newpage
% ===========================
\section{Appendix}

A more formal description is given as appendix.

% ===========================
\section{A1. Background} 
\label{sec:background}

Any meaningful motion of hand is called a hand \emph{gesture}.
Data captured during a gesture is called \emph{gesture data}.
In this study, gesture data is captured by means of an accelerometer
while user doing her hand gesture.
Therefore, the gesture data is a sequence of
an acceleration vectors in 3D.

% ===========================
\subsection{Notation}

Index $\delta = 1, 2, 3$ is exclusively used for the first, the second and 
the third dimensions of 3D. 
The $\delta^{\text{th}}$ component of 3D column vector $\vect{v}$ 
is denoted by $[\vect{v}]^{\delta}$.  

Acceleration vector 
$\vect{a} \triangleq [ a^{1}, a^{2}, a^{3} ]^{\top}$ 
is an 3D column vector where 
$\triangleq$ is used for definitions, and
$\top$ is the transpose operator.
An acceleration vector sampled at discrete time $k$ is represented as 
$\vect{a}(k) \triangleq [ a^{1}(k), a^{2}(k),$ $ a^{3}(k) ]^{\top}$.
A sequence of acceleration vectors sampled at discrete times 
$k = 1, 2, \dotsc, K$ is represented as 
\[
	\vect{A} \triangleq \{ \vect{a}(k) \}_{k=1}^{K}
\]
and called a \emph{gesture data}. 
Note that $\vect{A}$ is a 3D sequence.

Let $\mathcal{M} = \{1, 2, \dotsc, M\}$ be the set of $M = 20$ gestures.
Index $m \in \mathcal{M}$ is exclusively used to represent a gesture.
$\mathcal{A}_{m}$ represents the set of all gesture data for gesture $m$.
Then, $\mathcal{A} = \cup_{m=1}^{M} \mathcal{A}_{m}$ is the set of all gesture data.

Let 
$\vect{A}_{j} = \{ \vect{a}_{j}(k) \}_{k=1}^{K_{j}} \in \mathcal{A}_{m}$ be a gesture data of gesture $m$.
The average of $\vect{A}_{j}$ is defined as
\[
	\overline{\vect{A}_{j}} 
	= [\overline{a_{j}^{1}}, \overline{a_{j}^{2}}, \overline{a_{j}^{3}}]^{\top} 
	\triangleq
	\frac{1}{K_{j}} 
	\sum_{k=1}^{K_{j}} \vect{a}_{j}(k).
\]
Note that 
$\overline{a_{j}^{\delta}} = \left[ \ \overline{\vect{A}_{j}} \ \right]^{\delta}$ 
is the average of 
$\{ a_{j}^{\delta}(k) \}_{k=1}^{K_{j}}$ 
in dimension $\delta$.
Similarly, the average of all the gesture data in 
$\mathcal{A}_{m}$ is defined as 
\[
	\overline{\mathcal{A}_{m}} 
	\triangleq
	\frac{1}{|\mathcal{A}_{m}|} 
	\sum_{\vect{A}_{j} \in \mathcal{A}_{m}} \overline{\vect{A}_{j}}
\]
where $|\mathcal{A}_{m}|$ is the number elements in $\mathcal{A}_{m}$.
Note that both $\overline{\vect{A}_{j}} $ and 
$\overline{\vect{\mathcal{A}}_{m}}$ are 3D vectors.

Let $X$ be a gesture data to be classified. 
The true class of $X$ is denoted by $m_{t}(X)$
where as 
$m_{c}(X)$ is the class where the classifier assigns to.

% ===========================
\subsection{Template Matching Classification} 

Consider $M$-classes of 1-D sequences.
Let the sequence $R_{m} = \{r_{m}(i)\}_{i=1}^{I}$, called \emph{template}, 
be the representative of class $m$. 
We use a \emph{template matching classifier} 
which classifies sequence 
$X = \{x(j)\}_{j=1}^{J}$ 
to the class which minimizes distance of $X$ to its template~\cite{Alpaydin2010}.

In this work dynamic time warping cost is used as the distance
which calls for warping window $W$~\cite{Ratanamahatana2004}.
The quality of the classifier is improved by changing the warping window 
$W_{m}$ for each class $m$ rather than one $W$ for all.
A final remark is needed before the dynamic time warping technique
used in this study is elaborated.
Dynamic time warping is defined for 1-D signals. 
We extend this to 3D by simple summation of distances in each dimension
as given in \refeq{eq:DTWin3D}.

% ===========================
\subsection{Dynamic Time Warping (DTW)} 

Dynamic Time Warping (DTW) is an algorithm for finding the optimal match 
between two given time series
which may vary in time or speed with some certain 
criteria~\cite{Sakoe1978, Myers1981}. 
DTW is also used for measuring the similarity distance between two sequences 
after finding optimal match. 
Essentially DTW is in 1-D.

% ===========================
\subsubsection{Matching Cost}

Let $X = \{ x(i) \}_{i=1}^{I}$ and $Y = \{ y(j) \}_{j=1}^{J}$ be 
two sequences of real numbers with length $I$ and $J$, respectively.
Note that $X$ and $Y$ are time series in 1-D and of different lengths.
Then the DTW distance of $X$ and $Y$ is 
\[
	\DTW(X, Y) \triangleq c_{I, J}
\]
where $c_{I, J}$ is the $I^{\text{th}}$ and $J^{\text{th}}$ entry of an 
$I \times J$ matching cost matrix $\vect{C}$.
The entries of $\vect{C}$  are recursively defined as
\begin{align}
	c_{i, j} &\triangleq \gamma_{i, j}
	\label{eq:DTW}
\end{align}
for 
$i \in \{1, 2, \dotsc, I \}$ and
$j \in \{1, 2, \dotsc, J \}$
where 
\[
	\gamma_{i, j} 
	\triangleq
	|x(i) - y(j)| 
	+ \min \{
		c_{i, j-1}, \;
		c_{i-1, j-1}, \;
		c_{i-1, j} 
	\}
\]
with boundary conditions
\begin{align*}
	c_{0, 0} &= 0, \\
	\{c_{i, 0}\}_{i=1}^{I} &= \infty, \\
	\{c_{0, j}\}_{j=1}^{J} &= \infty.
\end{align*}

% ===========================
\subsubsection{Lower-Bound Keogh (LBK)}

The time and space complexity of $\DTW(X, Y)$ is $O(IJ)$
where $I$ and $J$ are the lengths of $X$ and $Y$, respectively.
In order to avoid unnecessary DTW calculation,
a lower bounding technique, called \emph{Lower-Bound Keogh (LBK)}, 
proposed by Keogh~\cite{Keogh2004}.
In template matching classifier, 
we need to find the closest template $R_{m}$ to $X$.
The minimum distance is found by iteratively calculating the distances to the templates and getting the minimum.
Suppose $s$ is the shortest distance so far.
Then instead of calculating the distance $\DTW(X,R_{j})$ to 
the $j^{\text{th}}$ template,
calculate much faster lower bound $\LBK(X,R_{j})$.
Since $\LBK \leq \DTW$, 
there is no need to calculate $\DTW(X,R_{j})$ 
as long as 
$s \leq \LBK(X,R_{j})$.
Only for cases 
$\LBK(X,R_{j}) < s$,
$\DTW(X,R_{j})$ is calculated.
If $\DTW(X,R_{j}) < s$,
then $s$ is updated with the new lower value, 
that is, $\DTW(X,R_{j})$.

% ===========================
\subsubsection{Warping Window} 

Given $X$ and $Y$, DTW produces a distance $\DTW(X, Y)$ 
but this may not be an accepted match.
It is possible that expanding or shrinking goes too far that 
corresponds to the cases where the matching path could be too far away 
from the diagonal~\cite{Sakoe1978}.
This is a well studied problem.
In order avoid matching paths going too away from the diagonal 
some restrictions in the form of \emph{warping window} are 
introduced~\cite{Sakoe1978, Ratanamahatana2004}.
Let $W = \{ w(k) \}_{k=1}^{\max\{I,J\}}$ be an adjustment window. 
Then the points with $w(n)$ away from diagonal should not be used, 
that is, \refeq{eq:DTW} is revised as 
\[
c_{i, j} 
\triangleq 
\begin{cases}
	\infty, & |i-j| \geq w(\max\{i,j\}), \\
	\gamma_{i, j,} & \text{otherwise}.
\end{cases}
\]
Then the distance of $X$ to $Y$ using warping window $W$ is 
denoted by $\DTW(X, Y, W)$.

% ===========================
\subsubsection{Ratanamahatana-Keogh Band (RK-Band)}

The shape of window $W$ needs to be decided according to application. 
One way to decide on $W$ is 
Ratanamaha--tana-Keogh Band (RK-Band) proposed in \cite{Ratanamahatana2004}.
In RK-Band, $W$ is iteratively changed to
optimize some criterion.

In our case the criterion is a metric of classification defined as follows.
Let $\mathcal{Y}$ be a set of gestures.
Elements $Y_{j} \in \mathcal{Y}$ are classified.
The total distances from $Y_{j}$ to the templates for correct and incorrect classification are defined as 
\[
	D_{c} \triangleq \sum_{
	\begin{subarray}{c}
		Y_{j} \in \mathcal{Y}\\
		m_{c}(Y_{j}) = m_{t}(Y_{j})
	\end{subarray}
	} \DTW(Y_{j}, R_{m_{c}(Y_{j})}, W)
\]
and
\[
	D_{i} \triangleq \sum_{
	\begin{subarray}{c}
		Y_{j} \in \mathcal{Y}\\
		m_{c}(Y_{j}) \neq m_{t}(Y_{j})
	\end{subarray}
	} \DTW(Y_{j}, R_{m_{c}(Y_{j})}, W),
\]
respectively.
The number of correct and incorrect classifications are
\[
	N_{c} \triangleq \sum_{
	\begin{subarray}{c}
		Y_{j} \in \mathcal{Y}\\
		m_{c}(Y_{j}) = m_{t}(Y_{j})
	\end{subarray}
	} 1
	\text{ \; and \; }
	N_{i} \triangleq \sum_{
	\begin{subarray}{c}
		Y_{j} \in \mathcal{Y}\\
		m_{c}(Y_{j}) \neq m_{t}(Y_{j})
	\end{subarray}
	} 1,
\]
respectively. 
Then, use the quality metric of \cite{Ratanamahatana2004} defined as
\begin{align}
	Q 
	\triangleq 
	\frac{D_{c}  N_{i}}{D_{i}  N_{c}}.
	\label{eq:Q}
\end{align}
Note that the value of $Q$ increases with a wrong classification and 
decreases with a correct classification.
More than that it is weighted with the distance.

% ===========================
\section{A2. Data Processing Blocks}
\label{sec:DataProcessingBlocks}

Raw gesture data 
$\vect{A} = \{ \vect{a}(k) \}_{k=1}^{K}$ 
which is collected from users 
is passed 
in some operations. 
The operations are represented as processing blocks.
Since the gesture data is in 3D,
the 1-D techniques given in Section~\ref{sec:background} 
need to be modified for 3D.
For example, the DTW distance in 3D is defined to be the summations of 
the individual DTW distances in each dimension $\delta$, that is,
\begin{equation}
	\DTW(\vect{X}, \vect{Y}, \vect{W})
	\triangleq
	\sum_{\delta=1}^{3}  
		\DTW( [\vect{X}]^{\delta}, [\vect{Y}]^{\delta}, [\vect{W}]^{\delta}). 
	\label{eq:DTWin3D}
\end{equation}

% ===========================
\subsection{Validation}

Clearly, every user has her own paste of doing a gesture. 
Some does the gesture fast, some does it slow.
Similarly, some user does the same gesture in a small scale, 
some in a large scale.
We discard gesture data that is too short or too long in duration,
i.e., 
$K < K_{min} \triangleq 30$ or 
$K_{max} \triangleq 205 < K$. 
The average amplitude of a gesture 
$\vect{A} = \{ \vect{a}(k) \}_{k=1}^{K}$ is defined as 
$a_{avg} \triangleq \frac{1}{K} \sum_{k=1}^{K} |\!| \vect{a}(k) |\!|$
where 
$|\!| \vect{a}(k) |\!|$ is the magnitude of 
$\vect{a}(k)$.
Data sets that are too small or too big in average amplitude,
that is, 
$a_{avg} < A_{min} \triangleq 0.95$ or 
$A_{max} \triangleq 2.10 < a_{avg}$, 
are also discarded.
Out of 1,090 data sets, 
24 due to duration and 4 due to amplitude, 
in total 28 are discarded and 
we end up with 1,062 gesture data for 20 gesture classes.

% ===========================
\subsection{Low-pass Filter}

The high frequency components are removed by means of a low-pass filter given as
$y_{k} = \alpha x_{k} + (1 - \alpha) y_{k-1}$
where
$x$ and $y$ are the input and the output signals of the filter, respectively and
$\alpha$ is the smoothing factor taking to be $\alpha=1/7$.
From now on, $\vect{A}_{j}$ means the low-pass filtered version of 
raw gesture data $\vect{A}_{j}$.

% ===========================
\subsection{Adjustment of Mean and Variance}

We adjust the mean and variance of all low-pass filtered gestures 
$\vect{A}_{j}$ in the set go gesture data
$\mathcal{A}_{m}$ for each class $m$ individually.
The gesture data 
$\vect{B}_{j} = \{ \vect{b}_{j}(k) \}_{k=1}^{K_{j}}$ 
with adjusted mean is obtained as
\[
	b_{j}^{\delta}(k) 
	\triangleq 
	a_{j}^{\delta}(k) 
	- \left[ \ \overline{\vect{A}_{j}} \ \right]^{\delta}
	+ \left[ \ \overline{\mathcal{A}_{m}} \ \right]^{\delta}
\]
with 
$k=1, 2, \dotsc, K_{j}$.
Let $\vect{v}_{j} = [ v_{j}^{1}, v_{j}^{2}, v_{j}^{3} ]^{\top}$ 
be the variance vector of $\vect{A}_{j}$
where
\[
	v_{j}^{\delta} 
	\triangleq 
	\frac{1}{K_{j}} 
	\sum_{k=1}^{K_{j}}
	\left( a_{j}^{\delta}(k) - \overline{a_{j}^{\delta}} \right)^{2}.
\]
Then the average variance of $\mathcal{A}_{m}$ would be
\[
	\overline{\vect{v}_{m}} 
	\triangleq 
	\frac{1}{|\mathcal{A}_{m}|} 
	\sum_{A_{j} \in \mathcal{A}_{m}} \vect{v}_{j}.
\]
Finally,
transform all gesture data $A_{j}$ in $\mathcal{A}_{m}$ to 
both mean and variance modified ones represented by
$\vect{C}_{j} = \{\vect{c}_{j}(k)\}_{k=1}^{K_{j}}$
where
\[
	c_{j}^{\delta}(k)
	\triangleq 
	\overline{b_{j}^{\delta}} + 
	\sqrt{\frac{\overline{v_{m}^{\delta}}}{v_{j}^{\delta}}} 
	\cdot \left( b_{j}^{\delta}(k) -  \overline{b_{j}^{\delta}} \right).
\]

% ===========================
\subsection{Down Sampling}

So far each gesture data has different duration. 
We down sample each gesture data in such a way that 
they have the same durations of $N \triangleq 30$,
which is the acceptable minimum duration as $K_{min}=30$.
Let $\{C_{j}(k)\}_{k=1}^{K_{j}}$ be the mean and 
average adjusted gesture data with duration, $K_{j}$.
Then the down sampled gesture data 
$\vect{D}_{j} = \{\vect{d}_{j}(n)\}_{n=1}^{N}$ 
is obtained by
\[
	\vect{d}_{j}(n) 
	\triangleq 
	\frac{1}{\Delta} 
	\sum_{\substack{k, \\ (n-1) \Delta < k \leq n \Delta}} 
		\vect{c}_{j}(k)
\]
for $n = 1, 2, \dotsc, N$
where
$\Delta \triangleq K_{j}/N$ is the \emph{downsampling factor}.

% ===========================
\subsection{Templates}

For each gesture $m$, we want to generate a template 
$\vect{G}_{m}$ so that 
a given gesture data 
$\vect{X}$ is classified to 
class $m_{j}$ if $\vect{X}$ is closest to $\vect{G}_{m_{j}}$ with respect to a distance metric.
The set of templates is denoted by 
$\mathcal{G} \triangleq \{ \vect{G}_{1}, \vect{G}_{2}, \dotsc,\vect{G}_{M}\}$

The template 
$\vect{G}_{m} = \{ \vect{g}_{m}(n) \}_{n=1}^{N}$ 
of class $m$ 
is obtained by averaging all the gesture data of the gesture $m$
as
\[
	\vect{g}_{m}(n) 
	\triangleq \frac{1}{|\mathcal{A}_{m}|}
	\sum_{A_{j} \in \mathcal{A}_{m}} \vect{d}_{j} (n).
\]

Besides templates $\vect{G}_{m}$, template generation also produces 
lower $\vect{L}_{m}$ and 
upper $\vect{U}_{m}$ bounds for each gesture $m$.
During classification, DTW is used as the distance metric.
In order to speed up,
the $\LBK$ technique is employed which requires 
$\vect{L}_{m}$ and 
$\vect{U}_{m}$ of each gesture class $m$.
$\vect{L}_{m}$ and $\vect{U}_{m}$ are calculated in two steps:
(i)~The lower bound $\vect{L}_{j}$ and upper bound $\vect{U}_{j}$ of 
$\vect{A}_{j}$ in the gesture set $\mathcal{A}_{m}$ is calculated 
for each dimension $\delta$ individually as given in \cite{Keogh2004} 
using $\LBK$ parameter $r=3$.
(ii)~Then, the lower bounds of the gesture set is obtained by averaging,
that is:
\[
	\vect{l}_{m} (n) 
	\triangleq \frac{1}{|\mathcal{A}_{m}|}
	\sum_{\vect{A}_{j} \in \mathcal{A}_{m}} \vect{l}_{j}(n).
\]
where $n = 1, 2, \dotsc, N$
for $\vect{L}_{m} = \{ \vect{l}_{m} (n) \}$.
For upper bounds, $\vect{U}_{m}$ are defined similarly.
Note that $\vect{l}_{j}(n)$ and $\vect{u}_{j}(n)$ are all 3D vectors.

% =========================== 
\subsection{Warping Window Size}

For each gesture class $m$, 
a specific sequence of warping window sizes 
$\vect{W}_{m} = \{ \vect{w}_{m}(n) \}_{n=1}^{N}$ 
is generated where $\vect{w}_{m}(n)$ is the window size at time $n$.
Warping window size generation is based on Ratanamahatana and Keogh's 
work~\cite{Keogh2004}. 
The warping window size $w(n)$ minimizes  
the quality metric $Q$ given in \refeq{eq:Q}, that is,
\[
	w(n) 
	= \underset{w}{\arg\max} \quad 
		\{ Q \}
\]
at each step $n \in \{ 1, 2, \dotsc, N \}$.

% ===========================
\subsection{Threshold Values}

Consider the distances of  $\vect{A}_{j} \in \mathcal{A}_{m}$ 
to template 
$\vect{G}_{m}$.
The minimum 
%$\vect{\Phi}^{min} = \{ \phi^{min}_{m_{m}}  \} _{m=1}^{M}$ and 
and maximum 
%$\vect{\Phi}^{max} = \{ \phi^{max}_{m_{m}}  \} _{m=1}^{M}$
of these distances are given as
\[
	\phi^{min}_{m}
	\triangleq 
	(1 - K_{\Phi})
	\min_{\vect{A}_{j} \in \mathcal{A}_{m}}
	\{ 
		\DTW( \vect{A}_{j}, \vect{G}_{m}, \vect{W}_{m}) 
	\} 
\]
and
\[
	\phi^{max}_{m}
	\triangleq 
	(1 + K_{\Phi})
	\max_{\vect{A}_{j} \in \mathcal{A}_{m}}
	\{ 
		\DTW( \vect{A}_{j}, \vect{G}_{m}, \vect{W}_{m})
	\}, 
\]
respectively,
where $K_{\Phi}$ is a safety constant taken to be $K_{\Phi} \triangleq 0.1$.

% ===========================
\subsection{DTW Template Matching}

Gesture $\vect{A}_{j}$ is classified to gesture class $m_{c}$ if 
$\DTW( \vect{A}_{j}$, $\vect{G}_{m_{c}}$, $\vect{W}_{m_{c}}) $ 
is the smallest for all $m \in \mathcal{M}$. 
That is
\[
	m_{c} 
	= \underset{m}{\arg\max} \quad 
		\{ 
			\DTW( \vect{A}_{j}, \vect{G}_{m}, \vect{W}_{m} ) 
		\}.
\]

This calls for repeated evaluation of 
$\DTW(\vect{A}_{j}, \vect{G}_{m}$, $\vect{W}_{m})$ for each $m$. 
The evaluation is speeded up by means of $\LBK$ technique 
using $L_{m}$ and $U_{m}$ generated in the template generation.

% ===========================
\subsection{Threshold Control}

Threshold values generated previously for given gesture class is used 
for classification result validation.
If 
$\phi^{min}_{m_{c}} 
< \DTW(\vect{A}_{j}, \vect{G}_{m_{c}}, \vect{W}_{m_{c}}) 
< \phi^{max}_{m_{c}} $,
then $m_{c}$ is the valid classification result. 
Otherwise, $m_{c}$ is discarded and classification result is invalid.

% =========================== 
\bibliographystyle{abbrv}
%\bibliographystyle{spbasic}

% =========================== 
\bibliography{DogukanCalculator}

% =========================== 
\end{document}